\documentclass[preprint,sort&compress]{elsarticle}

\usepackage{epsfig}
\usepackage{subfigure}
\usepackage{wrapfig}
\usepackage{multirow}
\usepackage{graphicx}

\begin{document}

\begin{frontmatter}

\title{Massively Parallel Computing and the Search for Jets and Black Holes at the LHC}

\author{V. Halyo\corref{cor1}}
\ead{vhalyo@gmail.com}
\author{P. LeGresley}
\author{P. Lujan}
\address{Department of Physics, Princeton University, Princeton, NJ 08544, USA}

\cortext[cor1]{Corresponding Author}

\begin{abstract}
Massively parallel computing at the LHC could be the next leap necessary to reach an era of new discoveries at
the LHC after the Higgs discovery. Scientific computing is a critical component of the LHC experiment, including
operation, trigger, LHC computing GRID, simulation, and analysis. One way to improve the physics reach of the LHC is to
take advantage of the flexibility of the trigger system by integrating coprocessors based on Graphics Processing Units 
(GPUs) or the Many Integrated Core (MIC) architecture into its server farm. This cutting edge technology provides not 
only the means to accelerate existing algorithms, but also the opportunity to develop new algorithms that select events
 in the trigger that previously would have evaded detection. In this article we describe new algorithms that would allow 
to select in the trigger new topological signatures that include non-prompt jet and black hole--like objects 
in the silicon tracker.
\end{abstract}


\end{frontmatter}

\section{Introduction} 

Data analysis at the LHC requires approaches to processing that are effective at finding the underlying physics phenomena while still being cost efficient.  Using computing technologies derived from consumer products results in lower hardware acquisition costs because many of the research, development, and manufacturing costs are amortized over much larger volumes.  These consumer products are increasingly mobile devices such as phones, tables, and notebook computers so there is significant interest in energy efficiency to maximize battery life.  This focus on energy efficiency also has the potential to benefit more traditional areas of computing and data analysis by reducing operating costs such as electrical power and cooling.

Leveraging processors derived from consumer products helps minimize costs associated with purchasing and operating large computing installations such as those required for the LHC, but there is also a software aspect that needs to be considered.  Moore's Law continues to yield ever more powerful processors, but the architecture of processors has changed significantly over the past decade.  For many years, CPUs were able to run the exact same software faster and faster with each new processor generation.  This changed in the early 2000s as Intel and AMD shifted focus to multicore processors featuring at first two cores, but with some newer processors featuring as many as 16 cores.

Parallel processors, in the form of multicore CPUs and more recently highly programmable Graphics Processing Units (GPUs), may require a significant rethink of algorithms and their implementations.  At the same time the High Energy Physics (HEP) community is at a crossroads with tentative confirmation of the Higgs particle completing the search for particles predicted by the Standard Model.  The search for Beyond the Standard Model (BSM) physics will require enhancement of the LHC trigger with parallel processors and computational accelerators such as GPUs or Xeon-Phi. In parallel it would require the development of advanced algorithms for detecting rare new physics phenomena. These new algorithms will need to be designed and implemented based on knowledge of the current state and likely future directions for processor architecture to allow and extend the physics reach at the LHC like never before.

\section{Physics Motivation}

Various BSM extensions predict the existence of new, strongly interacting particles that lead to final states
with high jet multiplicities~\cite{bib:sixjets,bib:multijets,bib:tcgut,bib:exoquarks,bib:unichiral,bib:altchiral}. Other exotic models might include boosted 
jets~\cite{Thaler:2008ju,Altheimer:2012mn} or long-lived neutral particles decaying at macroscopic distances 
from the primary vertex~\cite{bib:hiddenvalley,Halyo:2013yfa} 
in the tracker. The key to observing these events is selecting these jets in real time in the trigger system for archiving
 and further careful, offline analysis.

Both CMS~\cite{Chatrchyan:2008aa} and ATLAS~\cite{Aad:2008zzm} have a typical trigger system with a hierarchy of multiple levels, 
ranging from fast and relatively simple criteria implemented entirely in hardware and firmware, to more sophisticated software-based analysis. 
The primary goal of the software based level is to apply a specific set of 
physics selection algorithms on the events read out and accept the events with the most interesting physics 
content. This computationally intensive processing is executed on a farm of commercial CPU processors.
The quest for new physics and the flexibility of the trigger system suggests that this computer farm is the natural
place to integrate GPU or MIC cards. These cards allow for the development of fast and unique algorithms using massively
 parallel programming architectures to select events that would indicate BSM physics. 

In this article newly developed trigger algorithms are described that would be able to select events that
include displaced jets originating from long lived particles, or various topological signatures with boosted
jets that will be more frequent as the center of mass energy of the collision at design luminosity is
increased. These new triggers will naturally be sensitive to a larger parameter space, including lower-mass
Higgs that decay to displaced heavy flavor hadrons, that could have evaded detection in previous
experiments. For example, CMS has performed a search for long-lived neutral particles decaying to displaced
leptons~\cite{Chatrchyan:2012jna}, but the efficiency for measuring leptons originating from a low-mass Higgs
of $M_H = 120$ GeV/$c^2$ is very low due to the lepton momentum requirements in the trigger, resulting in
relatively poor limits for this mass region.

Generalization of the trigger algorithm developed in this article would also allow searching for 
the existence of soft displaced black hole--like decay; that is, high multiplicity omnidirectional decay of soft particles  
that originate a few centimeters from the interaction point. To conclude, introducing massively parallel programming
allows for development of algorithms that are computationally infeasible using CPUs and
might provide an invaluable way to test for topological signatures that clearly do not exist in the SM in real time.

\section{Processor Architecture}

When discussing computer processors it is common to bring up Moore's Law.  However, it is important to recognize what Moore's Law actually represents.  Moore's Law is an empirical observation made by Intel co-founder Gordon Moore in the 1960s that it was economically feasible to double the number of transistors on a single integrated circuit every 24 months,
emphasizing the manufacturing technology and the ability to fabricate ever smaller transistors.

In terms of performance, one of the most obvious benefits of smaller transistors is the ability to run processors at faster clock rates.  Over the course of a few years processor clock speeds jumped from tens of MHz to GHz speeds.  However, in recent years power and heat restrictions have limited the further increase of processor clock speed as a means for obtaining additional power from an existing architecture.

It is fairly clear how a faster clock speed yields a faster processor.  What is a little less obvious is how more transistors can make a single core processor faster even at fixed clock speeds.  In the late 1990s Intel and AMD added to their processors SSE, or Streaming SIMD Extensions, where Single Instruction Multiple Data (SIMD)
refers to a type of parallel architecture as defined by Flynn's taxonomy.  In the SIMD  model one instruction such as multiplication or addition is applied to multiple data items.  This parallelism is typically something to be expressed by the software developer based on the computations to be performed and the capabilities of the hardware.  However, most modern compilers can also target the SSE units in a process typically referred to as auto-vectorization.  The effectiveness of auto-vectorization depends on the capabilities of the compiler, the complexity of the code being analyzed, and the skill of the software developer at writing code that can be successfully analyzed by the compiler.

SSE requires specific software instructions to make use of it, either through manual programming by the software developer or auto-vectorization at compile time.  The additional transistors available at each generation of Moore's Law have also allowed hardware designers to implement Instruction Level Parallelism (ILP).  Instead of the processor executing one operation per clock cycle, multiple instructions are executed in parallel. The implementation of ILP and the associated techniques of instruction pipelining, out-of-order execution, speculative execution, and hardware branch prediction helped consume many of the additional transistors made available through Moore's Law.  But in addition to ILP, the growing gap between the performance of the processors running at GHz speeds and the off-chip memory systems necessitated adding a complicated cache hierarchy to modern processors.

\begin{figure}[!Hhtb]
\begin{center}
\includegraphics[width=0.5\textwidth]{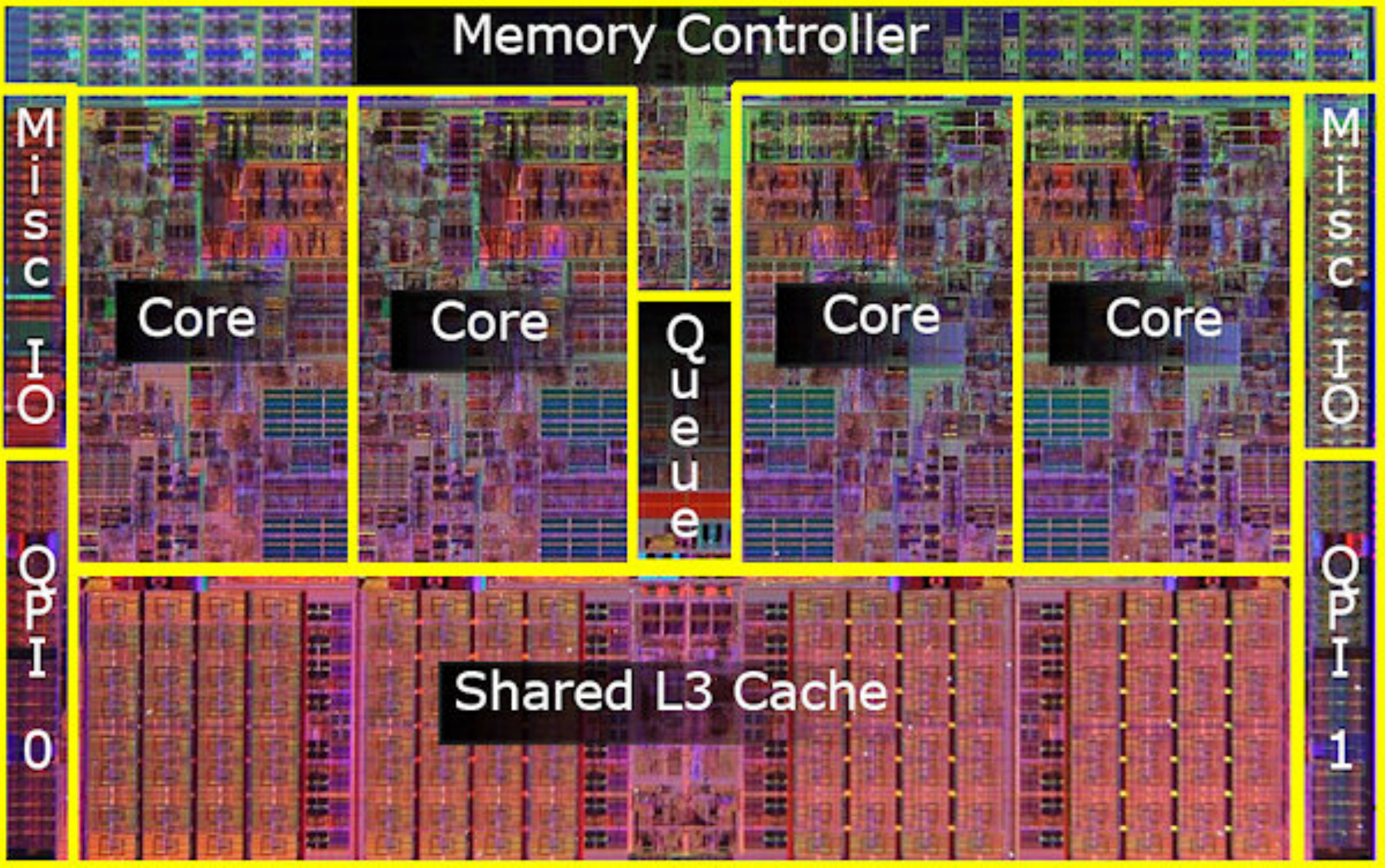}
\caption{Intel Core i7 die showing major components of the processor (image from Intel). \label{fig:core_i7_die}}
\end{center}
\end{figure}

While computing performance may increase by around 40\% per year in accordance with Moore's Law, memory performance increases approximately 10\% per year.  If left unchecked this gap in performance would result in processors that had the potential to be very computationally powerful but would be limited in actual performance because of the constant need to wait on the memory system.  To address this, modern processors have a complicated cache hierarchy not only for data, but also for instructions, and even for translating virtual memory addresses to physical memory addresses in the form of Translation Lookaside Buffers (TLB).  Additional logic in the form of hardware prefetchers observes memory access patterns in real time and attempts to prefetch into the caches data and instructions that are likely to be needed in the future.  The disparity between processor and memory performance is so significant that the vast majority of the chip area in modern processors is devoted not to computation but to data handling.  As shown in Figure~\ref{fig:core_i7_die} the L3 cache plus the memory controller and other I/O portions of a chip can easily represent around 50\% of the area.  Within each core there is an L1 and L2 cache, plus all of the logic associated with ILP for that core.  The actual chip area devoted to computations is remarkably small.

Faster clock speeds, ILP, and hardware prefetchers have the advantage of using existing software and simply running it faster.  However, there are limits to what the hardware can do to automatically extract ILP and predict the memory access pattern of arbitrary applications.  Deep instruction pipelines, which in some processors have exceeded 30 stages, have issues with hazards that would affect the correctness of results if not correctly handled by the processor, stalls due to unavailable inputs, and mispredicted branching behavior.  At some point this reaches a level of diminishing returns where the hardware complexity, and hence cost, outweighs the performance benefits.

With faster clock speeds not feasible and ILP also reaching its limits, hardware designers in the early 2000s turned to multicore processing to utilize all of those additional transistors available to them from Moore's Law.  As seen in Figure~\ref{fig:core_i7_die}, the hardware designers develop the design for a core, plus the hardware logic to link them together, and can then place as many cores on a single chip as current manufacturing technology allows.  With multiple cores, independent instances of applications can run in parallel on the multiple cores without application-level changes.  But if one application such as a video transcoder wants to speed up the processing of a single video, it must be specially written to make use of multiple cores.

Modern CPUs with ILP, SIMD units, and multicore are highly parallel processors, yet still retain the ability to efficiently run single-threaded, sequential software because so many applications make little to no use of parallel computing.  A different approach would be to design an inherently parallel processor that is only designed to run parallel programs efficiently.  Such a processor would not need to run single-threaded, sequential software and therefore could devote more chip area to computations and less to features like ILP, caches, and hardware prefetechers.

Graphics Processing Units (GPUs) are an example of a processor that is specialized to running parallel workloads, and a
recent trend is the use of systems consisting of processors with multiple cores integrated with GPUs as a modified form
of stream processor. This concept turns the massive floating-point computational power of a modern graphics
accelerator's shader pipeline into general-purpose computing power, as opposed to being hard-wired solely to do
graphical operations. In certain applications requiring massive vector operations, this can yield several orders of
magnitude higher performance than a conventional CPU.

In addition to making GPUs more computationally powerful, the programmability has been improved through both changes in hardware and new software tools.  The programmability of GPUs using options such as Compute Unified Device Architecture (CUDA), OpenCL, and OpenACC make the hardware usable for many applications beyond graphics, and without requiring programmers to have a specialized graphics background.  However, a strong background in parallel computing is still required to make effective use of a GPU.

With the growing success of GPUs as a massively parallel processor well suited to more general purpose applications, Intel has introduced a line of products based on a Many Integrated Core (MIC) architecture, marketed under the name Xeon Phi.  The Xeon Phi products physically look similar to a GPU in that they plug into a host system via PCI Express.  And from an architecture perspective, they also have many similarities to GPUs.  Xeon Phi is based on an x86 Pentium core architecture from the early 1990s.  This is a much simplified, and hence smaller, core compared to modern x86 CPUs.  To make these simplified cores more computationally powerful 512 bit wide SIMD units have been added to the core.  Because of the simplified nature of the core a single Xeon Phi coprocessor may have 60 or more cores each supporting 4 way Hyper-Threading (known more generally in computing as Simultaneous Multithreading (SMT)).

Regardless of whether the future is GPUs or the architecturally similar Xeon Phi, it seems clear that the future of high-performance numerical computing is massively parallel processors.  This will require rethinking algorithms and their implementations to make effective use of these highly parallel processors.  Algorithms that have been viewed as computationally infeasible should be revisited, and current algorithms that do not parallelize well will need to be reworked to take advantage of modern processors.

\section{Current Tracker and Tracking Algorithm}

\begin{figure}[!Hhtb]
\begin{center}
    \includegraphics[width=0.45\textwidth]{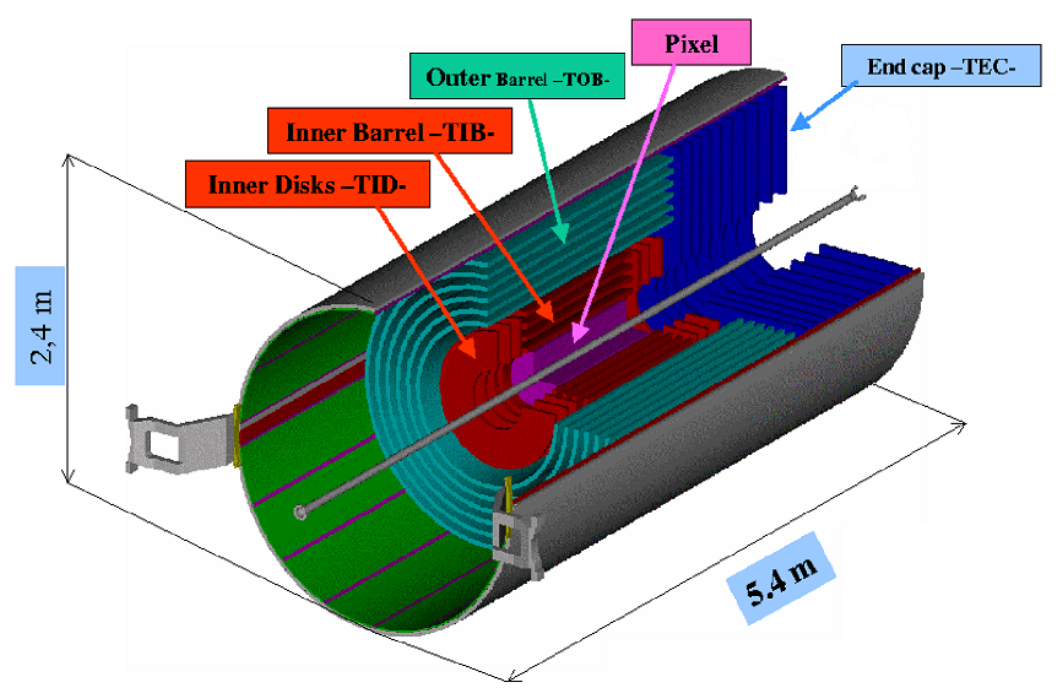}
    \includegraphics[width=0.45\textwidth]{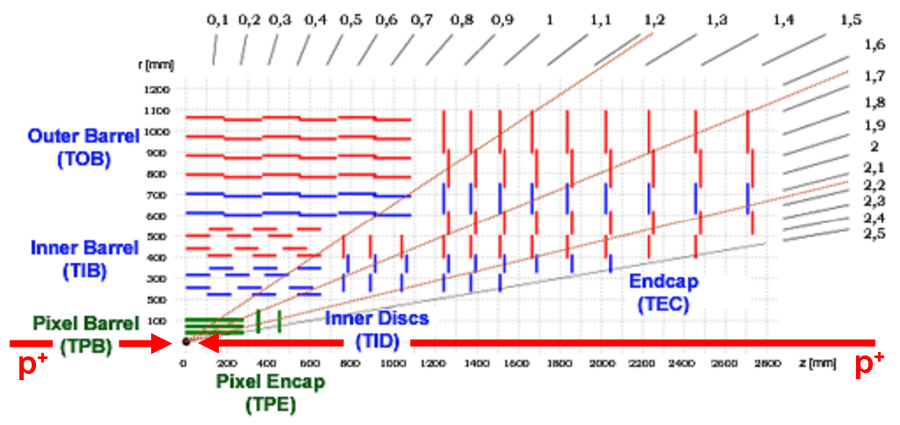}
    \caption{The CMS tracker in a 3-D view (left) and a 2-D view (right), showing the pixel detector and the silicon
    strip tracker. \label{fig:tracker}}
    \end{center}
\end{figure}

Both the CMS and ATLAS inner trackers include a silicon pixel detector and a silicon strip detector.
All tracker layers provide two-dimensional hit position measurements, but only the pixel tracker
 and a subset of the strip tracker layers provide three-dimensional hit position measurements.
Only the CMS tracker is considered; however, the detector performances are comparable
in the CMS and ATLAS experiments, and the tracking algorithm and results discussed are 
applicable to both of these experiments.
 The CMS pixel detector includes three barrel layers and two forward disks on either end of the
detector; lying outside the pixel detector is the strip detector, consisting of ten layers in the barrel plus
three inner disks and nine forward disks at each end of the detector. Owing to the strong magnetic field and
the high granularity of the silicon tracker, promptly-produced charged particles with transverse momentum $p_T
= 100$ GeV/c are reconstructed with a resolution in $p_T$ of 1.5\% and in transverse impact parameter $d_0$ of
15 mm. The track reconstruction algorithms are able to reconstruct displaced tracks with transverse impact
parameters up to 25 cm from particles decaying up to 50 cm from the beam line. The performance of the track
reconstruction algorithms has been studied with data in~\cite{bib:Khachatryan:2010pw}. The silicon tracker is also
used to reconstruct the primary vertex position with a precision of $\sigma_d = 20$ mm in each dimension.
Figure~\ref{fig:tracker} shows the layout of the CMS tracker.

The CMS track reconstruction, known as the Combinatorial Track Finder (CTF), uses an iterative algorithm, with
earlier iterations searching for tracks that are more easily found (relatively higher $p_T$ tracks close to
the interaction region). The hits in these tracks are then removed, allowing later iterations to search for
lower momentum or highly displaced tracks without the number of combinatorial possibilities becoming too
large. Each iteration consists of four steps: a seeding step, a track finding step, a track fitting step, and
a selection step.

In the first step, a seed, consisting of either three hits or two hits and a primary vertex constraint, is
constructed to create an initial estimate of the trajectory parameters for the track. Second, the track
finding is performed using a global Kalman filter~\cite{bib:Fruhwirth:1987fm}. This step performs a fast
propagation of the track candidate to propagate the track through the layers of the detector; at each layer, a
search for compatible nearby hits is performed, and these are attached to the candidate. After all candidates
in this step have been found, the track candidate collection is then cleaned to remove duplicate tracks or
tracks which share a large number of hits. The third step is to perform a full Kalman fit over the whole
track to obtain the best estimate of the track parameters at all points along the trajectory. The filter
begins at the innermost hits, and then iterates outward through each hit to update the track trajectory
estimate and its uncertainty. After this first fit is complete, a smoothing stage is then performed running
backwards from the last hit to apply the information from the later hits to the earlier ones. This step uses a
Runge-Kutta propagator to account for the effect of material interactions and an inhomogeneous magnetic
field. Finally, track selection requirements are applied to reduce the fake rate of the resulting track
candidates.

While the CTF is a well-established and reliable algorithm, its combinatorial nature means that as the LHC
luminosity, and hence the number of hits observed in the detector, increases, the number of possible
combinations increases at a faster rate. Thus, increases in hardware performance alone are not necessarily
enough to keep the running time of the algorithm reasonable as LHC luminosity continues to increase. This is a
particularly severe problem in the CMS trigger, where the available computation time is quite limited; as a
consequence, certain constraints on the tracks that can be reconstructed in the trigger are applied, to limit
the number of possible combinations. This limits our ability to search for new physics models which would
produce these kinds of tracks.

\section{Hough Transform Algorithm}

As a demonstration of how computationally intensive 2D tracking algorithms can take advantage of massively parallel GPU processing, the authors developed an implementation of the Hough transform using CUDA~\cite{Halyo:2013} in order to measure in real time the transverse momentum of prompt or non-prompt tracks.  The Hough transform~\cite{bib:HT1,bib:HT2,bib:HT3} is an image processing algorithm for feature detection that considers all possible instances of a parameterized feature such as a line or circle.  Each possible instance of a feature starts with zero votes in the parameter space, and then for each piece of input data votes are added to the feature instances that would include that input data.  After all input data has been processed the votes in the parameter space are processed.  Locations in the parameter space with more votes are likely to be actual features in the input data so this step amounts to looking for local maxima in the parameter space.  Once candidate features have been identified, more expensive computations can be applied to confirm the existence of the feature.

\begin{figure}[!Hhtb]
\begin{center}
  \subfigure[]{\includegraphics[width=0.32\textwidth]{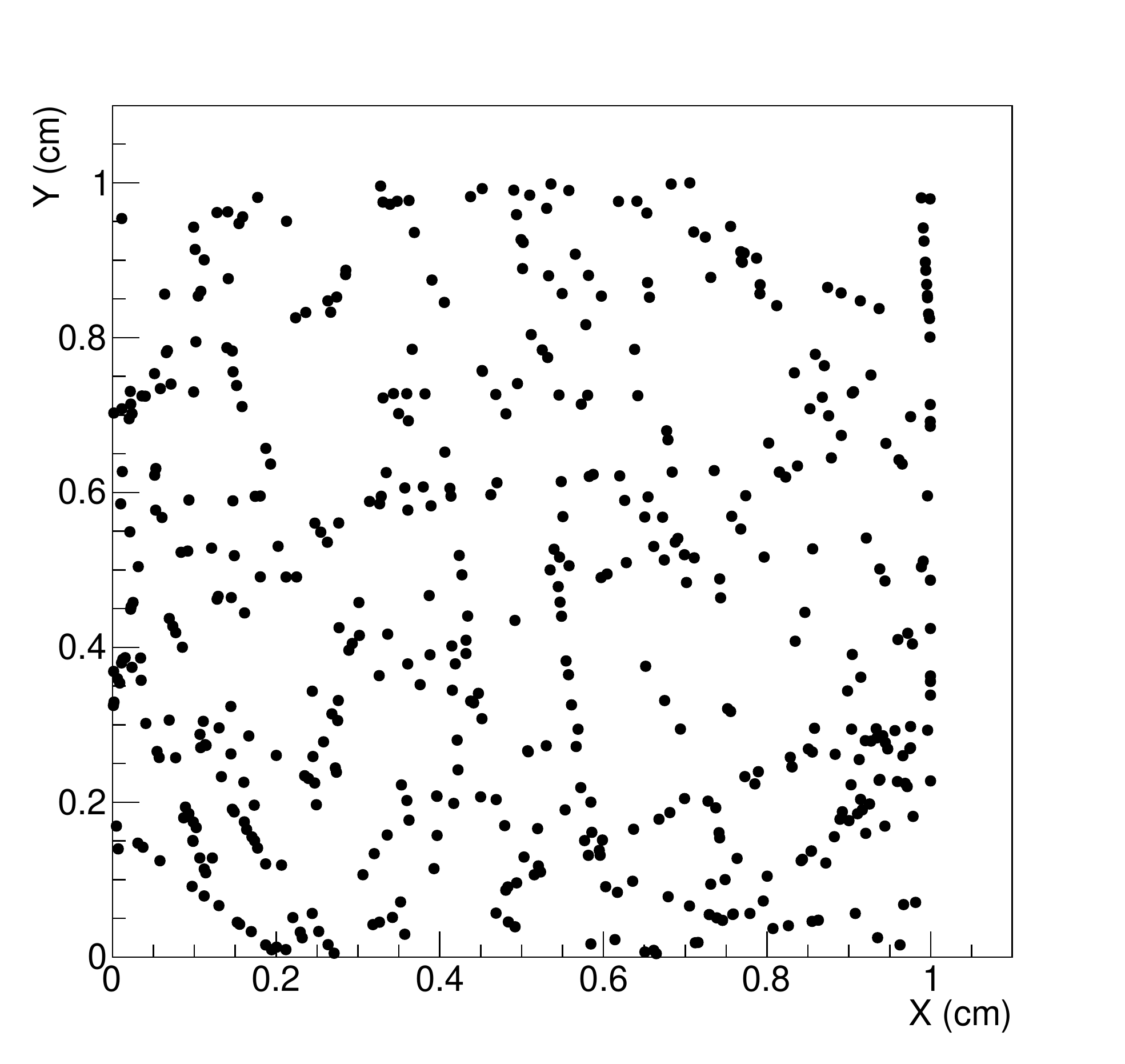} \label{fig:hits}}
  \subfigure[]{\includegraphics[width=0.30\textwidth]{50events_accumulator.pdf} \label{fig:accumulator}} 
  \subfigure[]{\includegraphics[width=0.32\textwidth]{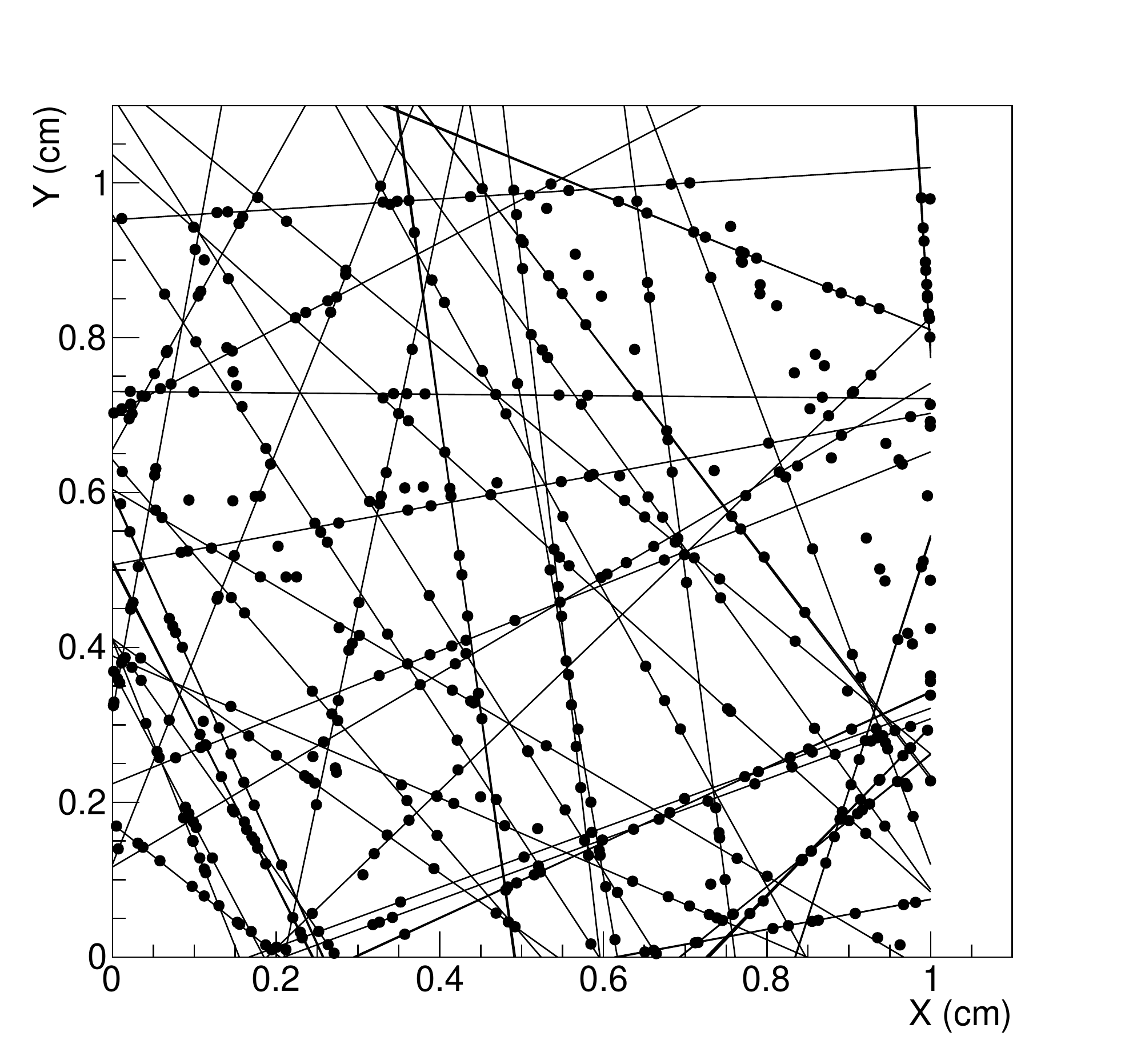} \label{fig:tracks}}
  \caption{Hough transform algorithm applied to a simple example. Left: Hits in a simulated event with 50
    straight-line tracks and 10 hits per track. Center: Each hit results in a sinusoidal curve of votes in parameter
    space. Locations with many votes are likely to be tracks in the original data. Right:
    Candidate tracks identified from finding local maxima in the parameter space.\label{fig:hough}}
\end{center}
\end{figure}

One advantage of the Hough transform is that it is tolerant to missing data or data that does not exactly fit the candidate features.  This could occur if there is noise in the data, or if the resolution of the data results in features that cannot be perfectly represented with a given computational discretization.  The downside of this robustness is that the Hough transform is computationally expensive.  However, the implementation using the GPU has demonstrated significant performance advantages compared to a multithreaded CPU implementation.  This performance comparison was made using a self-developed CPU implementation because typical implementations such as the one in the Intel Integrated Performance Primitives (IPP) library are sequential implementations that do not utilize multicore CPUs.  It was felt that the multithreaded CPU implementation was a more fair way to compare performance, but self-implementations can sometimes be biased.  Therefore, additional CPU performance optimizations have been implemented in the form of using Advanced Vector Extensions (AVX) intrinsics and improvements to the OpenMP parallelization.

Sample input data was generated using a Monte Carlo simulation of a simple detector model where only the transverse plane is considered.  The model contains a simulated beam pipe with a radius of 3.0 cm surrounded by ten concentric, evenly-spaced tracking layers with an overall radius of 110.0 cm.  A hit resolution of 0.4 mm in each direction is used.

\begin{figure}[!Hhtb]
\begin{center}
\includegraphics[width=0.75\textwidth]{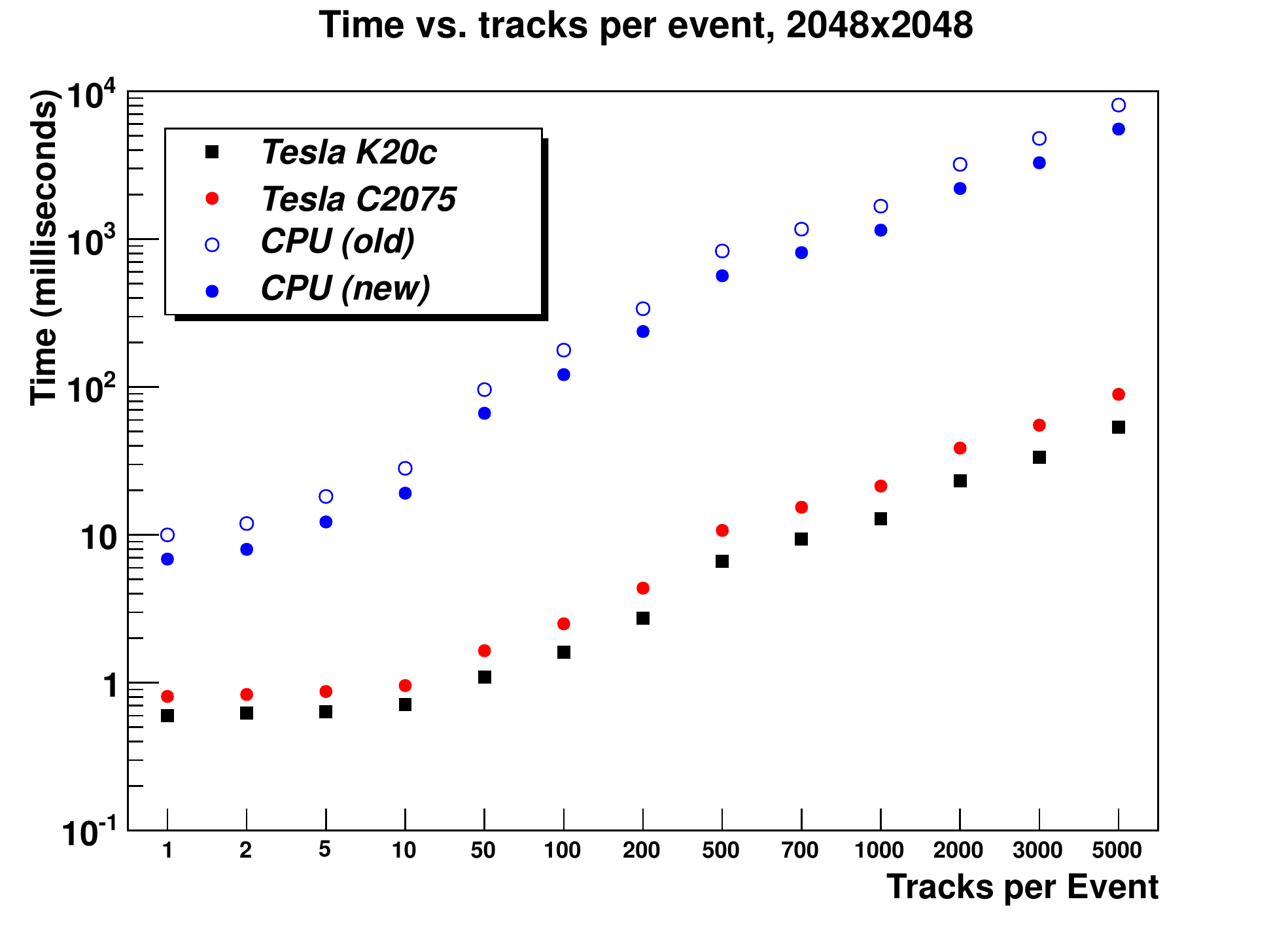} 
\caption{Time performance of old and new CPU implementations compared to NVIDIA Tesla C2075 and K20c GPUs. 
 Intel Core i7-3770 CPU running a multithreaded implementation.\label{fig:TimePerformance}}
\end{center}
\end{figure}

As shown in Figure~\ref{fig:TimePerformance} the combination of improvements resulted in approximately a 40\% decrease in the runtime for the CPU implementation.

It is worth mentioning that while the use of AVX has the potential for an 8x improvement when using single precision floating point, the actual observed performance improvement of 40\% was not unexpected for this code.  That is because the previous CPU implementation sought to minimize expensive computations by ensuring that redundant computations were moved outside of loops and intermediate results reused where possible.  These existing optimizations meant the implementation was limited more by memory access operations rather than the computational rate.  It is always useful to understand what aspect(s) of the hardware, such as computations or data access, limit the performance of an algorithm so that time spent on implementing optimizations is used effectively.

\section{Jet and Black Hole Detection}

\begin{figure}[!Hhtb]
\begin{center}
\includegraphics[width=0.75\textwidth]{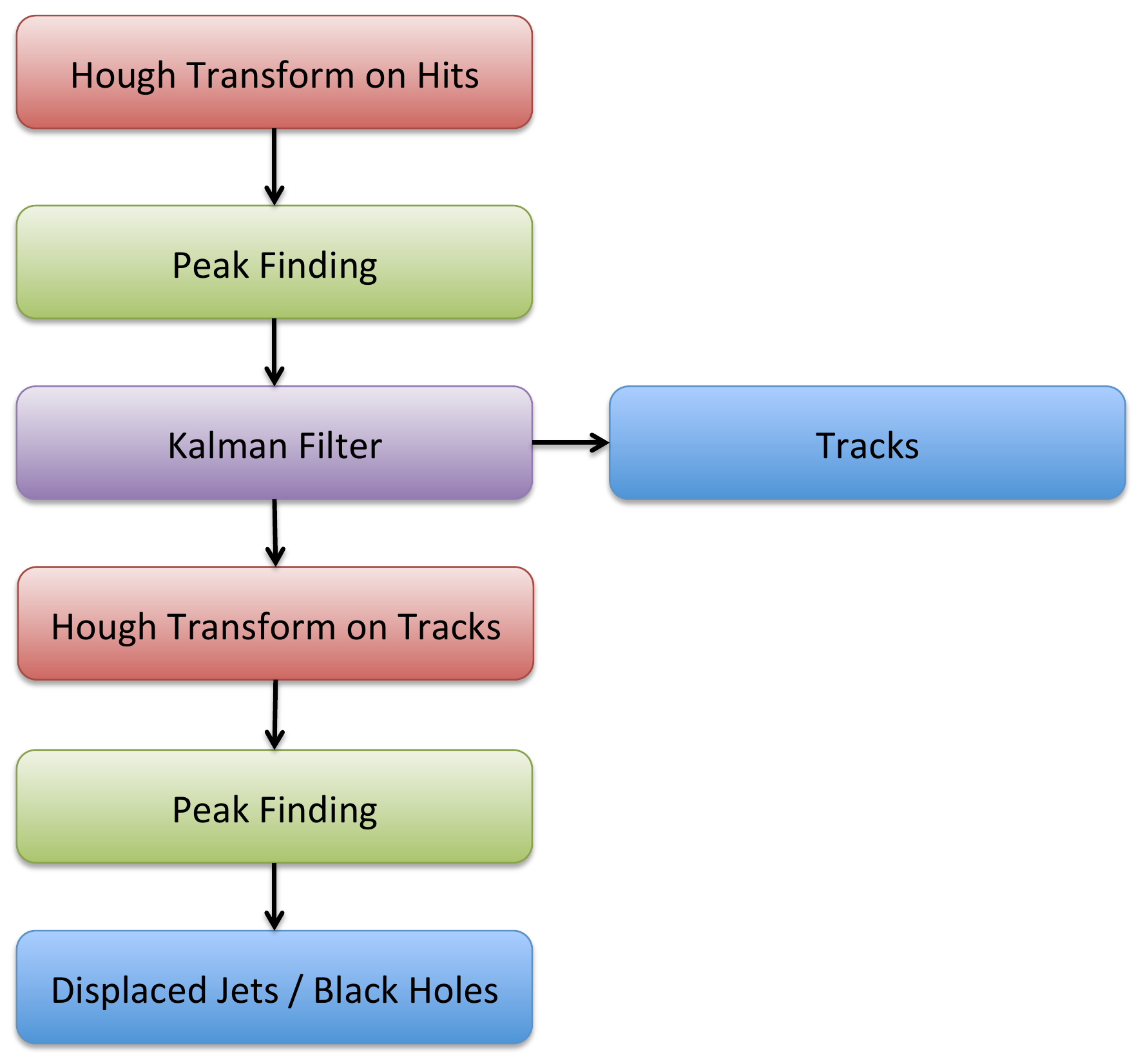} 
\caption{Flowchart showing the steps in processing the input hit data to tracks and displaced jets or black holes.\label{fig:Flowchart}}
\end{center}
\end{figure}

The overall data processing pipeline for detection of displaced jets and black holes builds on the existing components of the tracking algorithm as shown in Figure~\ref{fig:Flowchart}.  As an initial step the hit data for an event is processed using the Hough transform to compute the parameter space representation.  Candidate tracks are identified by finding local maxima, or peaks, in the parameter space.  A Kalman filter is then used to further refine the candidate tracks.  At this time the Kalman filter is being run using a CPU implementation due to the relatively insignificant amount of time required compared to other aspects of the data analysis.  The Kalman filter algorithm is amenable to a GPU implementation which could be undertaken if the computational cost of the Kalman filter becomes a bottleneck.

With candidate tracks identified the corresponding parameter space representation of these tracks is processed by a second application of the Hough transform.  In this case the goal is to identify not straight lines but sinusoids which contain a significant number of tracks, indicating that this is a location of a displaced jet or black hole.  Again, the location of features of interest correspond to local maxima, or peaks, in the parameter space so the same peak finding algorithm is applied to the parameter space of this second Hough transform.  The output of the peak finding can be used directly, as shown here, or if necessary after application of the Kalman filter.

It should be noted that the large number of tracks originating from the interaction point will also be identified as a jet.  These false positives can be excluded by filtering out features that are at or extremely close to the known interaction point.  This filtering could be done within the Hough transform process itself by setting a lower bound on the radius, or at the downstream peak finding step. The Kalman filter could also be incorporated at that step, if desired.

\begin{figure}[!Hhtb]
\begin{center}
	\includegraphics[width=0.35\textwidth]{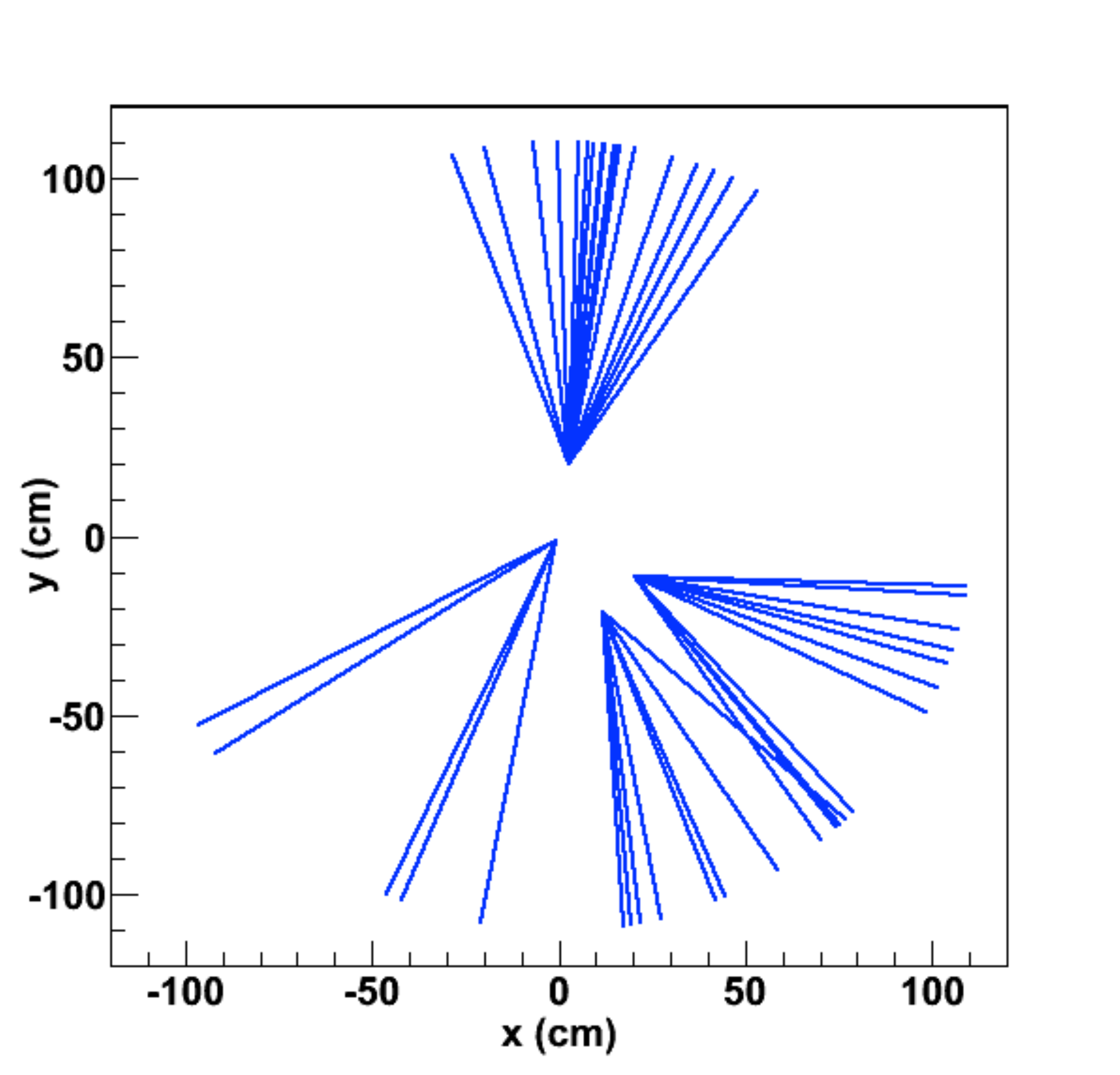}
	\includegraphics[width=0.30\textwidth]{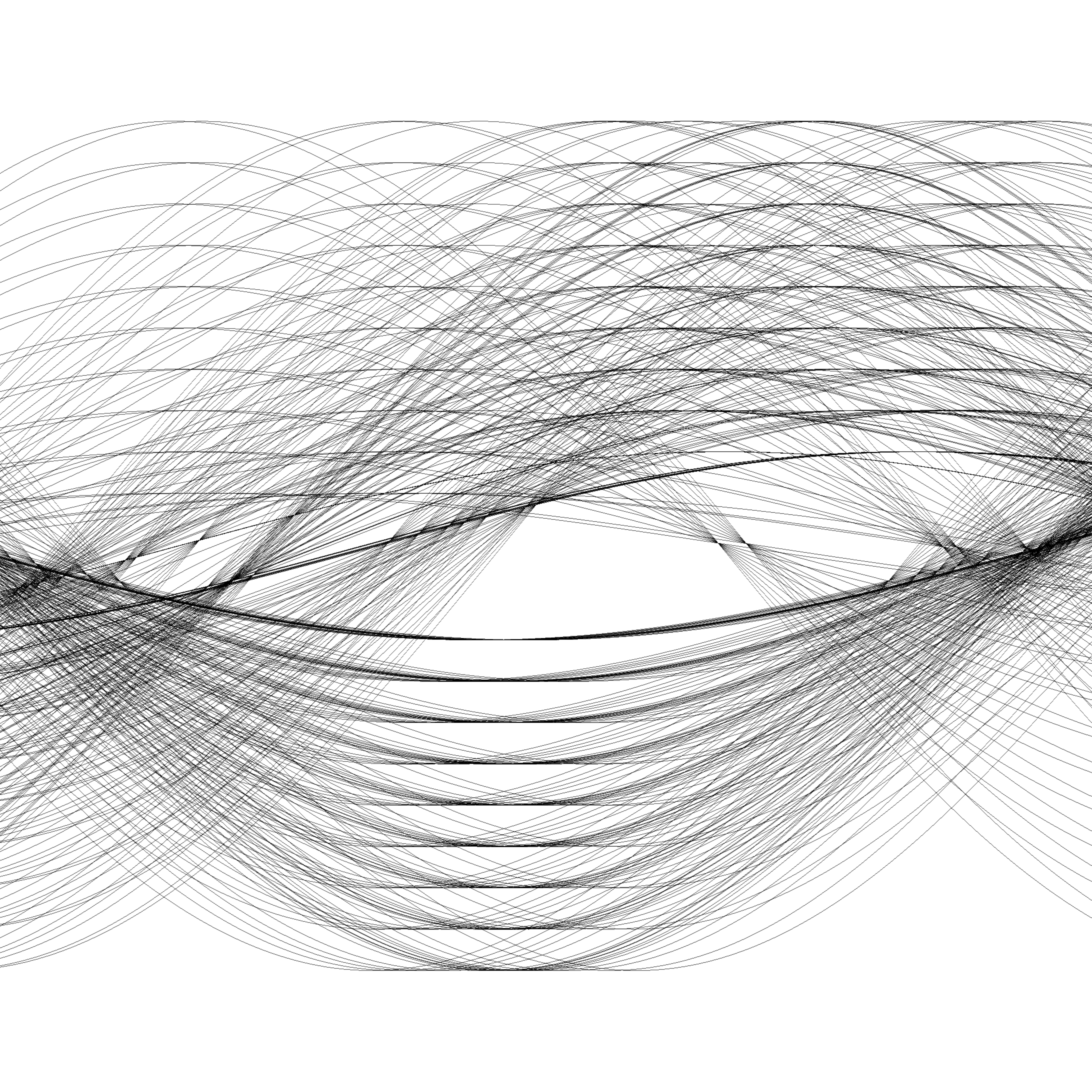}
	\includegraphics[width=0.30\textwidth]{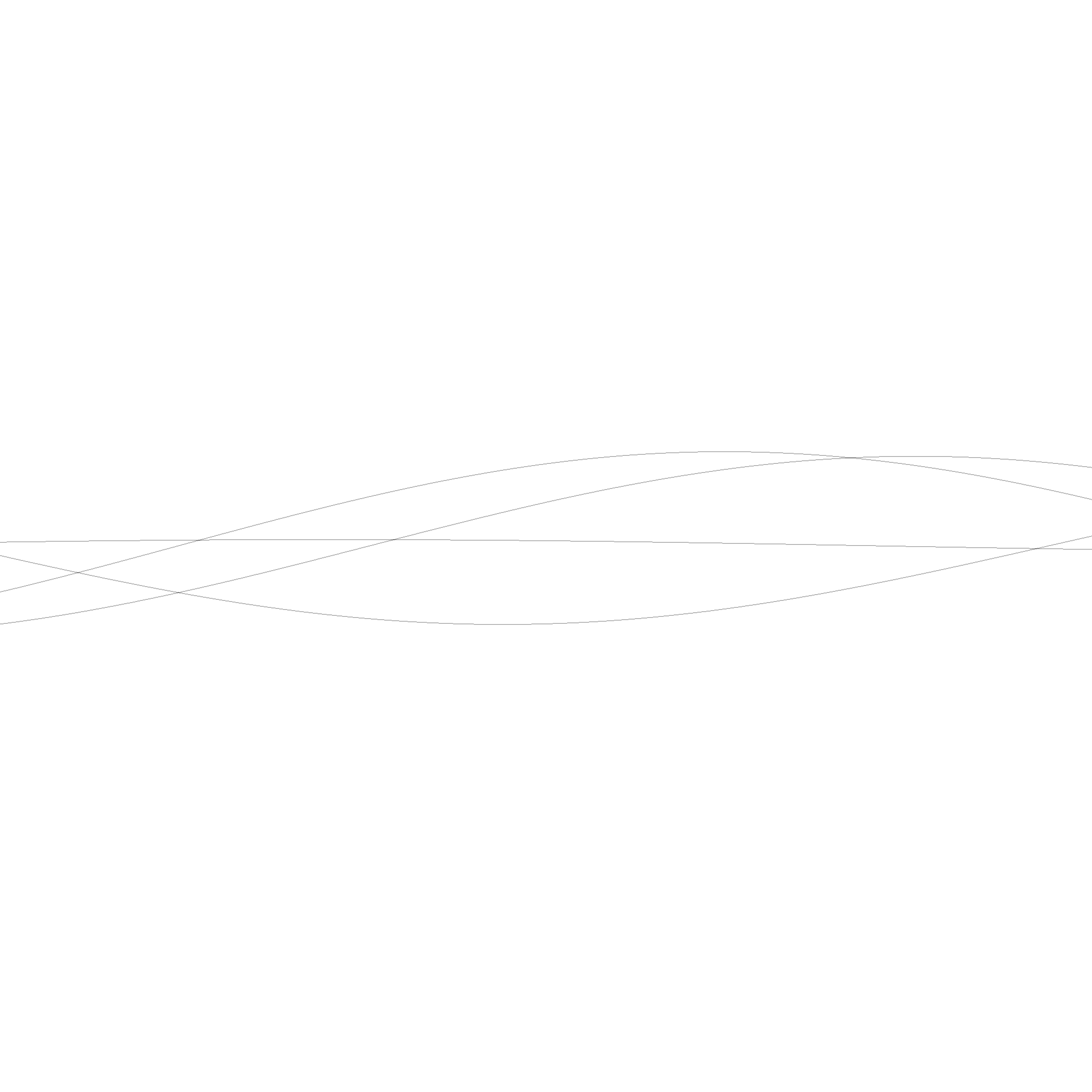}
	\caption{The Hough transform algorithm applied to an event with multiple displaced jets. Left: The simulated
	tracks present in the event. Center: The parameter space after applying the first Hough transform is still very
	cluttered. Right: The second Hough transform identifies the sinusoids corresponding to the jet vertices.
	\label{fig:DisplacedJets}}
\end{center}
\end{figure}

Figure~\ref{fig:DisplacedJets} illustrates the detection process for a simple event with four displaced jets.  The initial hit data is used to identify tracks as shown on the left in x--y space and in the center in parameter space.  Applying a second Hough transform to the tracks in parameter space identifies four sinusoids in the second parameter space shown on the right.  These four sinusoids correspond to the four vertices of the displaced jets in the original x--y space.

\section{Preliminary Results}

\begin{figure}[!Hhtb]
\begin{minipage}[t]{5.6cm}
\begin{center}
\includegraphics[width=1. \textwidth]{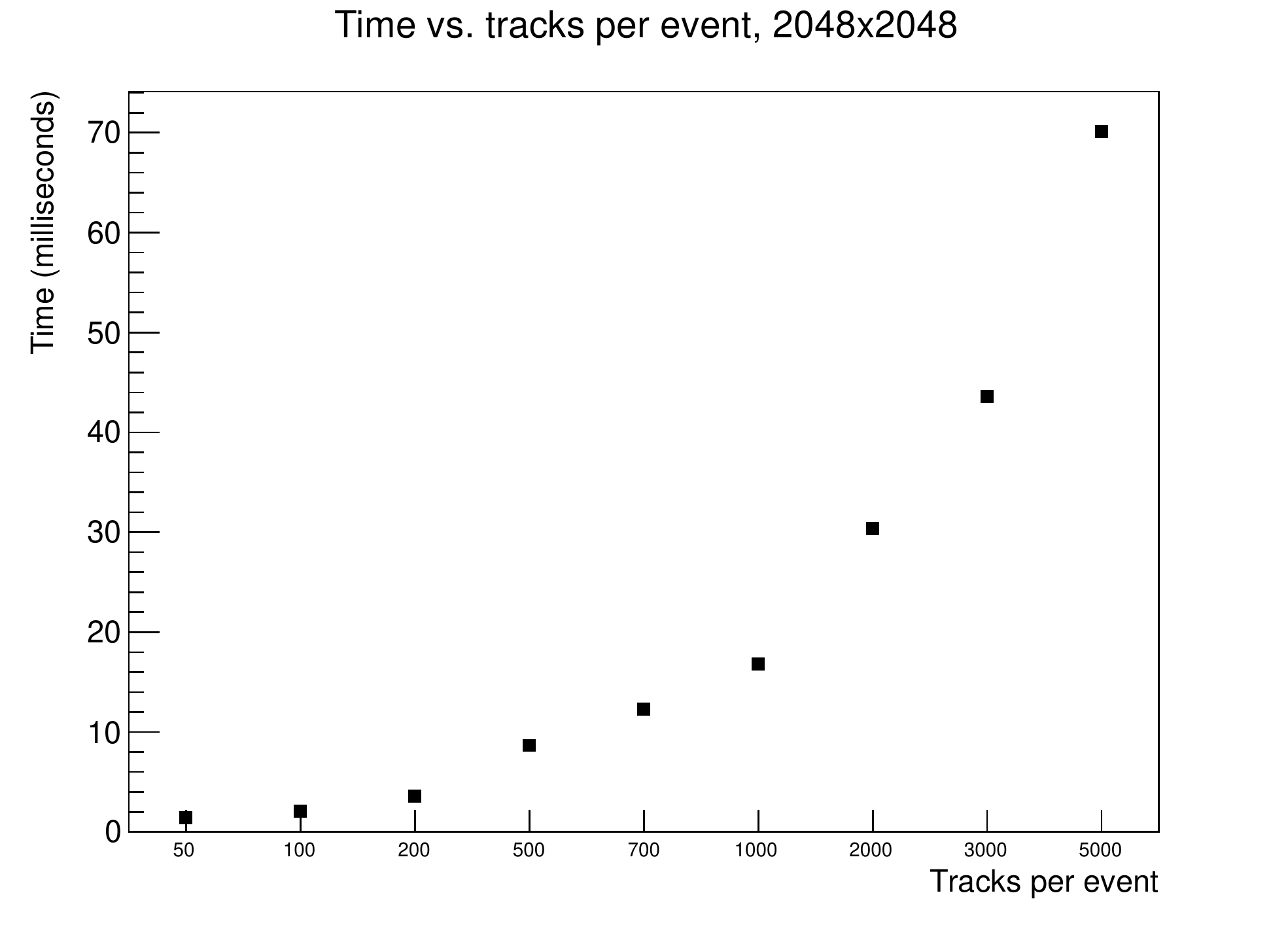}
 \caption{Time performance of the tracking plus displaced jet detection algorithm using an NVIDIA Tesla K20c GPU.  
          Ten percent of the tracks are associated with a displaced jet. \label{fig:performance}}
\end{center}
\end{minipage}\quad
\begin{minipage}[t]{5.7cm}
\begin{center}
\includegraphics[width=1.\textwidth]{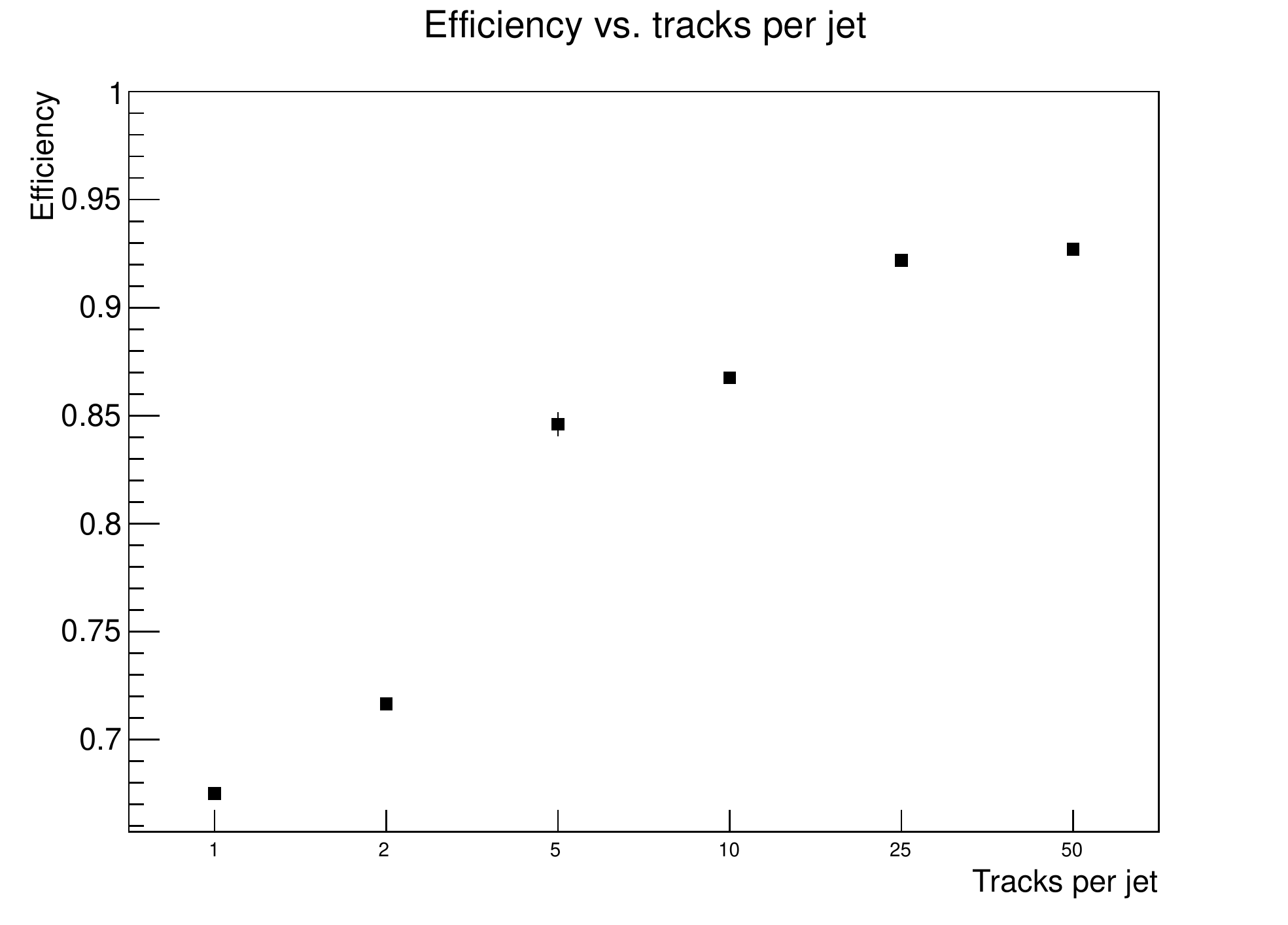} 
\caption{Efficiency of displaced jet detection with varying number of tracks per displaced jet.\label{fig:eff_num_tracks_per_singularity}}
\end{center}
\end{minipage}
\end{figure}

At first glance the proposed enhancement for detecting displaced jets and black holes would seem to double the computational cost due to the need to perform a second expensive Hough transform.  However, it is important to note that the first application of the Hough transform and associated peak finding and Kalman filtering results in a significant data dimensionality reduction.  Where the first application of the transform has a cost proportional to the number of hits, the second application of the transform has a cost proportional to the number of tracks.  For example, with an average of 10 hits per track, and in the limit of 100\% efficiency and purity, the dimensionality of the data is reduced by a factor of 10.  Figure~\ref{fig:performance} shows a time comparison of the baseline tracking algorithm and the enhancement for detecting displaced jets and black holes.  Ten percent of the tracks are associated with a displaced jet.  So, in practice, the computational cost increases by approximately 30\% compared to the baseline tracking algorithm.  This time includes the time for the data transfer and computational time of the Kalman filter running on the CPU and represents an additional opportunity for performance optimization if the Kalman filter were to be moved to the GPU.

In Figure~\ref{fig:eff_num_tracks_per_singularity} results for the efficiency of detecting a single jet with a varying number of tracks and 10 hits per track are shown.  Efficiency is defined as the fraction of simulated jets successfully identified by the algorithm divided by the number of jets known to be present.  The efficiency is relatively low for a small number of tracks but quickly rises for 10 tracks and beyond.  Because the peak finding has been optimized for track detection where 7 or more hits per track are expected, simple reuse of this code explains the efficiency results.  Separate tuning of the peak detection for a fewer number of votes per feature could improve the jet finding efficiency.

\begin{figure}[!Hhtb]
\begin{minipage}[t]{5.6cm}
\begin{center}
\includegraphics[width=1.\textwidth]{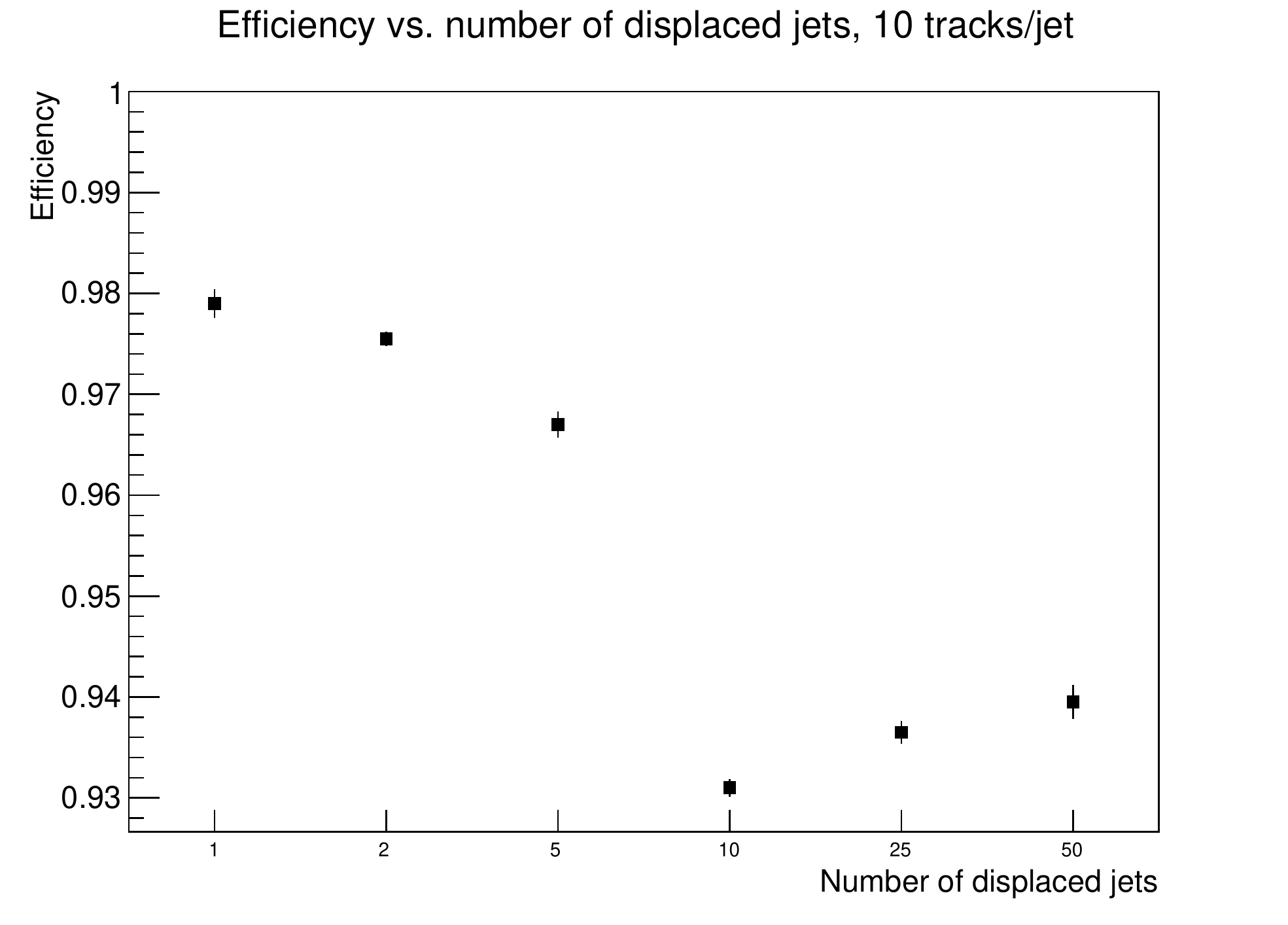} 
 \caption{Efficiency of displaced jet detection with varying number of displaced jets per event.  
Each displaced jet has 10 tracks and a dispersion angle between 15 and 60 degrees.  \label{fig:eff_num_jets}}
\end{center}
\end{minipage}\quad
\begin{minipage}[t]{5.7cm}
\begin{center}
\includegraphics[width=1.\textwidth]{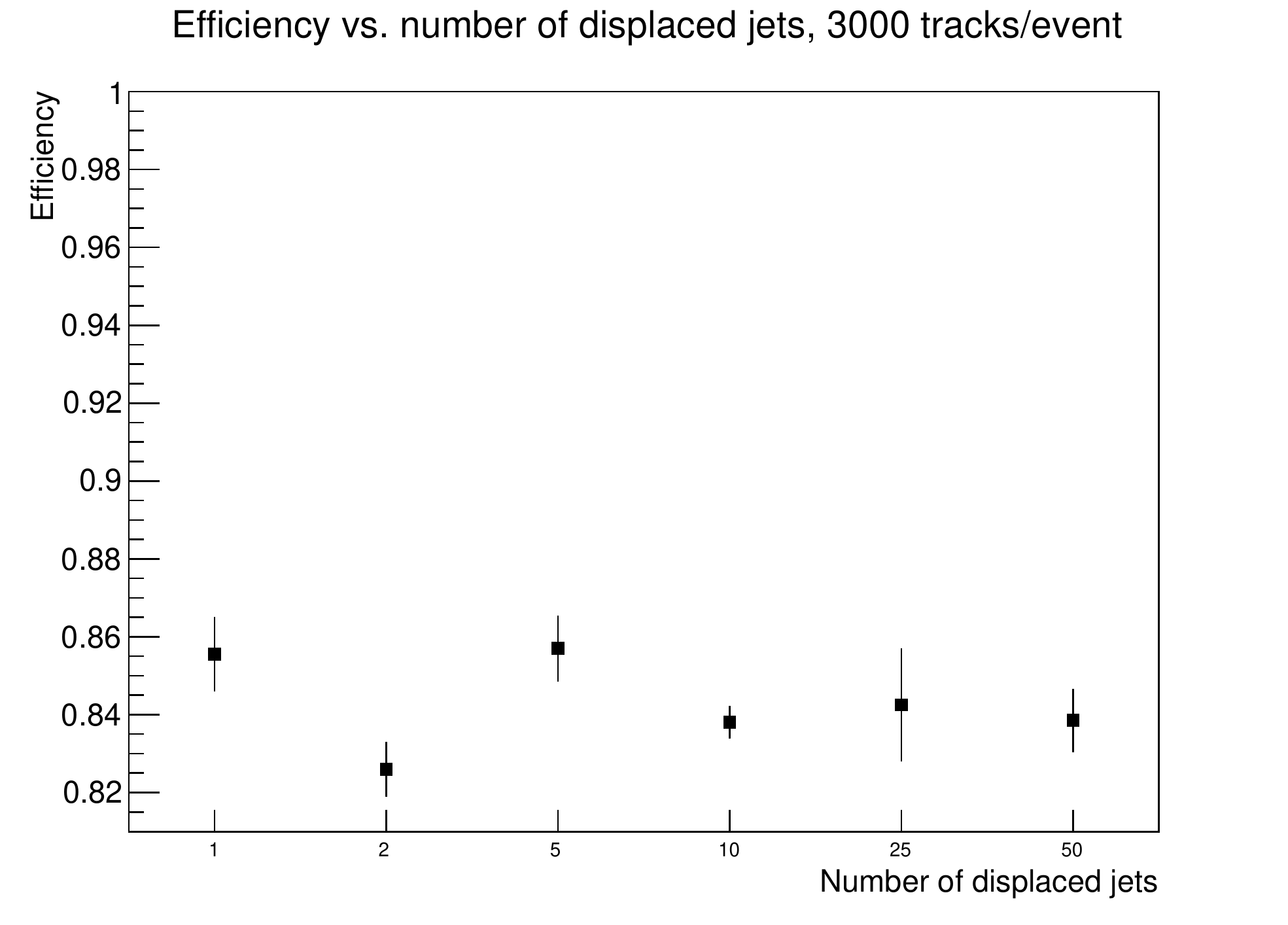}  
\caption{Efficiency of displaced jet detection in the presence of tracks originating from the interaction point.  
There are a total of 3000 tracks, some of which make up a varying number of displaced jets. \label{fig:eff_jets_plus_tracks}}
\end{center}
\end{minipage}
\end{figure}

Identification of a jet or black hole becomes more difficult if there are multiple such features in the event.  Figure~\ref{fig:eff_num_jets} shows efficiency with a varying number of jets per event.  Each jet has 10 tracks with 10 hits per track and dispersion angles between 15 and 60 degrees.  There seems to be a slight trend towards lower efficiency as the number of jets per event is increased.

Finally a more realistic event made up of 3000 total tracks of 10 hits each is considered, with some of the tracks being part of a varying number of jets made up of 10 tracks each and having dispersion angles between 15 and 60 degrees.  Because there are now tracks originating at the interaction point, the jet detection filters out any potential vertices within the 3.0 cm beam pipe radius.  As shown in Figure~\ref{fig:eff_jets_plus_tracks} the efficiency seems to be independent of the number of displaced jets.

\section{Summary}

Massively parallel computing is a critical component at the LHC experiment that is necessary in order to
extend its physics reach. Integration of coprocessors based on GPUs or MIC architecture into the LHC trigger
server farm provides not only the means to accelerate existing algorithms, but also the opportunity to develop
new algorithms that select events in the trigger that previously would have evaded detection.

A natural extension of a computationally intensive 2D tracking algorithm using the Hough Transform algorithm was developed to enhance the LHC trigger performance.
The GPU based algorithm has been extended to detect topological signatures that include displaced jets and black holes never possible before; 
the signature of these events would be a smoking gun for the presence of physics beyond the Standard Model. 
The proposed extension uses a second application of the Hough transform to identify tracks originating from a vertex displaced from the interaction point. 
 Although the Hough transform is a computationally expensive algorithm that originally could not have run in the trigger system, an implementation on a computational accelerator, such as a GPU, 
is significantly faster than implementations on conventional CPUs.  Recent work to incorporate additional CPU optimizations yielded approximately a 40\% improvement in performance, 
although the GPU implementation is still approximately 60x faster than the corresponding multithreaded, vectorized CPU implementation.  The computational cost increase when performing the second
 Hough transform is mitigated by the data reduction of the first application of the transform, which reduces the data dimensionality from the number of hits to the number of tracks. 
 Benchmarks using data from Monte Carlo simulations showed that the implementation of this proposed trigger enhancement for topological signatures that include 
displaced jets and black holes only increased the runtime by 30\%. Hence, the promising results of this trigger algorithms suggest that massively parallel computing at the LHC could be the next leap necessary 
to reach an era of new discoveries at the LHC post the Higgs discovery. 

Future work will focus on efficiency and purity improvements, as well as investigating further GPU performance optimization.  In addition, it would be desirable to compare the performance results presented in this article with an equivalent implementation on the Intel Xeon Phi~\cite{bib:Halyo:2013gja}.  While still based on the x86 architecture, the Xeon Phi is designed using much simpler cores, allowing for many more cores per processor like in GPUs.

To conclude, this work demonstrates one of the advantages of massively parallel computing at the LHC as a means
to reach an era of new discoveries at the LHC after the Higgs discovery.

\bibliographystyle{model1-num-names}
\bibliography{JetsBlackHoles}

\end{document}